\documentclass[11pt,a4paper]{article} 
\usepackage{jheppub,hyperref,mdwlist}
\usepackage{amsmath}
\usepackage{graphicx}
\usepackage{subfigure}
\usepackage{amssymb}
\usepackage[dvipsnames]{xcolor}
\usepackage{multirow}
\usepackage{cancel}
\usepackage{color}
\usepackage{listings}
\usepackage{xcolor}
\usepackage{float}
\usepackage{slashed}

\usepackage{tikz}
\usetikzlibrary{arrows,shapes}
\usetikzlibrary{trees}
\usetikzlibrary{matrix,arrows} 				
\usetikzlibrary{positioning}				
\usetikzlibrary{calc,through}				
\usetikzlibrary{decorations.pathreplacing}  
\usepackage{pgffor}							

\usetikzlibrary{decorations.pathmorphing}	
\usetikzlibrary{decorations.markings}
\tikzset{
    vector/.style={decorate, decoration={snake}, draw},
	provector/.style={decorate, decoration={snake,amplitude=2.5pt}, draw},
	antivector/.style={decorate, decoration={snake,amplitude=-2.5pt}, draw},
    fermion/.style={draw=black, postaction={decorate},
        decoration={markings,mark=at position .55 with {\arrow[draw=black]{>}}}},
    fermionbar/.style={draw=black, postaction={decorate},
        decoration={markings,mark=at position .55 with {\arrow[draw=black]{<}}}},
    fermionnoarrow/.style={draw=black},
    gluon/.style={decorate, draw=black,
        decoration={coil,amplitude=4pt, segment length=5pt}},
    scalar/.style={dashed,draw=black, postaction={decorate},
        decoration={markings,mark=at position .55 with {\arrow[draw=black]{>}}}},
    scalarbar/.style={dashed,draw=black, postaction={decorate},
        decoration={markings,mark=at position .55 with {\arrow[draw=black]{<}}}},
    scalarnoarrow/.style={dashed,draw=black},
    electron/.style={draw=black, postaction={decorate},
        decoration={markings,mark=at position .55 with {\arrow[draw=black]{>}}}},
	bigvector/.style={decorate, decoration={snake,amplitude=4pt}, draw},
}

\tikzstyle{block} = [draw, rectangle, 
    minimum height=3em, minimum width=6em]

\usepackage{amsmath}
\usepackage{wrapfig}
\usepackage{cleveref}

\newcommand{\be}{\begin{equation}}
\newcommand{\ee}{\end{equation}}
\newcommand{\beq}{\begin{equation}}
\newcommand{\eeq}{\end{equation}}
\newcommand{\bea}{\begin{eqnarray}}
\newcommand{\eea}{\end{eqnarray}}
\newcommand{\besp}{\begin{equation}\begin{split}}
\newcommand{\eesp}{\end{split}\end{equation}}

\newcommand{\Eq}[1]{Eq.~(\ref{#1})}

\newcommand{\Dfbd}{\mathord{\buildrel{\lower3pt\hbox{$\scriptscriptstyle\leftrightarrow$}}\over {D}_{\mu}}}
\newcommand{\ave}[1]{\left\langle #1\right\rangle}

\def\mL{\mathcal{L}}
\def\mM{\mathcal{M}}

\def\mO{\mathcal{O}}
\def\mP{\mathcal{P}}

\def\0{\textbf{0}}
\def\1{\textbf{1}}
\def\2{\textbf{2}}
\def\3{\textbf{3}}
\def\4{\textbf{4}}
\def\5{\textbf{5}}
\def\6{\textbf{6}}
\def\7{\textbf{7}}
\def\8{\textbf{8}}
\def\9{\textbf{9}}

\def\p{\textbf{p}}
\def\hc{\text{h.c.}}
\def\h{\mathfrak{h}}
\def\s{\mathfrak{s}}

\begin{document}

\title{Leptogenesis triggered by a first-order phase transition}

\author{Peisi Huang,}
\author{Ke-Pan Xie}
\affiliation{Department of Physics and Astronomy, University of Nebraska, Lincoln, NE 68588, USA}

\emailAdd{peisi.huang@unl.edu}
\emailAdd{kpxie666@163.com}

\abstract{
We propose a new scenario of leptogenesis, which is triggered by a first-order phase transition (FOPT). The right-handed neutrinos (RHNs) are massless in the old vacuum, while they acquire a mass in the new vacuum bubbles, and the mass gap is huge compared with the FOPT temperature. The ultra-relativistic bubble walls sweep the RHNs into the bubbles, where the RHNs experience fast decay and generate the lepton asymmetry, which is further converted to the baryon asymmetry of the Universe (BAU). Since the RHNs are out of equilibrium inside the bubble, the generated BAU does not suffer from the thermal bath washout. We first discuss the general feature of such a FOPT leptogenesis mechanism, and then realize it in an extended $B-L$ model. The gravitational waves from $U(1)_{B-L}$ breaking could be detected at the future interferometers.
}

\maketitle
\flushbottom

\section{Introduction}

Leptogenesis is a class of mechanisms for solving the matter-antimatter asymmetry problem of the Universe~\cite{Fukugita:1986hr,Luty:1992un,Davidson:2008bu}. In this paradigm, the heavy right-handed neutrinos (RHNs) $\nu_R$ decay to the Standard Model (SM) particles via the CP violating Dirac Yukawa interaction $\lambda_D\bar\ell_L\tilde H\nu_R$, generating a lepton asymmetry which is then converted to a baryon asymmetry of the Universe (BAU) via the electroweak (EW) sphaleron process. A particularly attractive and elegant feature of this paradigm is that the same coupling can also explain the origin of neutrino masses~\cite{Davis:1994jw,Super-Kamiokande:1998kpq,KamLAND:2002uet} via Type-I seesaw~\cite{Minkowski:1977sc}. In the conventional thermal leptogenesis scenario, the generated BAU is determined by the competition between the enhancement from the CP violating phase in the $\lambda_D$ matrix and the washout effects from the thermal bath. Typically, the washout processes are so efficient that only $\mO(10^{-2})$ of the originally generated BAU survives till today~\cite{Buchmuller:2005eh}. If different generations of RHNs have a mass hierarchy, then the CP violating phase has an upper bound proportional to the lightest $\nu_R$ mass (denoted as $M_1$), known as the Davidson-Ibarra bound~\cite{Davidson:2002qv}, which requires $M_1\gtrsim 10^9$ GeV to generate the observed BAU.\footnote{If the masses of at least two generations of $\nu_R$'s are nearly degenerate, the CP violating phase can be resonantly enhanced to be $\mO(1)$, independent of the $\nu_R$ mass. In that case, successful leptogenesis can occur for $\mO({\rm TeV})$ scale RHN~\cite{Flanz:1996fb,Pilaftsis:1997jf,Pilaftsis:2003gt,Dev:2017wwc}. If some of the RHNs do not reach thermal equilibrium before the EW sphaleron process is switched off, then leptogenesis may even apply to the sub-EW scale $\nu_R$~\cite{Akhmedov:1998qx,Drewes:2017zyw}.}

\begin{figure}
\centering
\includegraphics[scale=0.5]{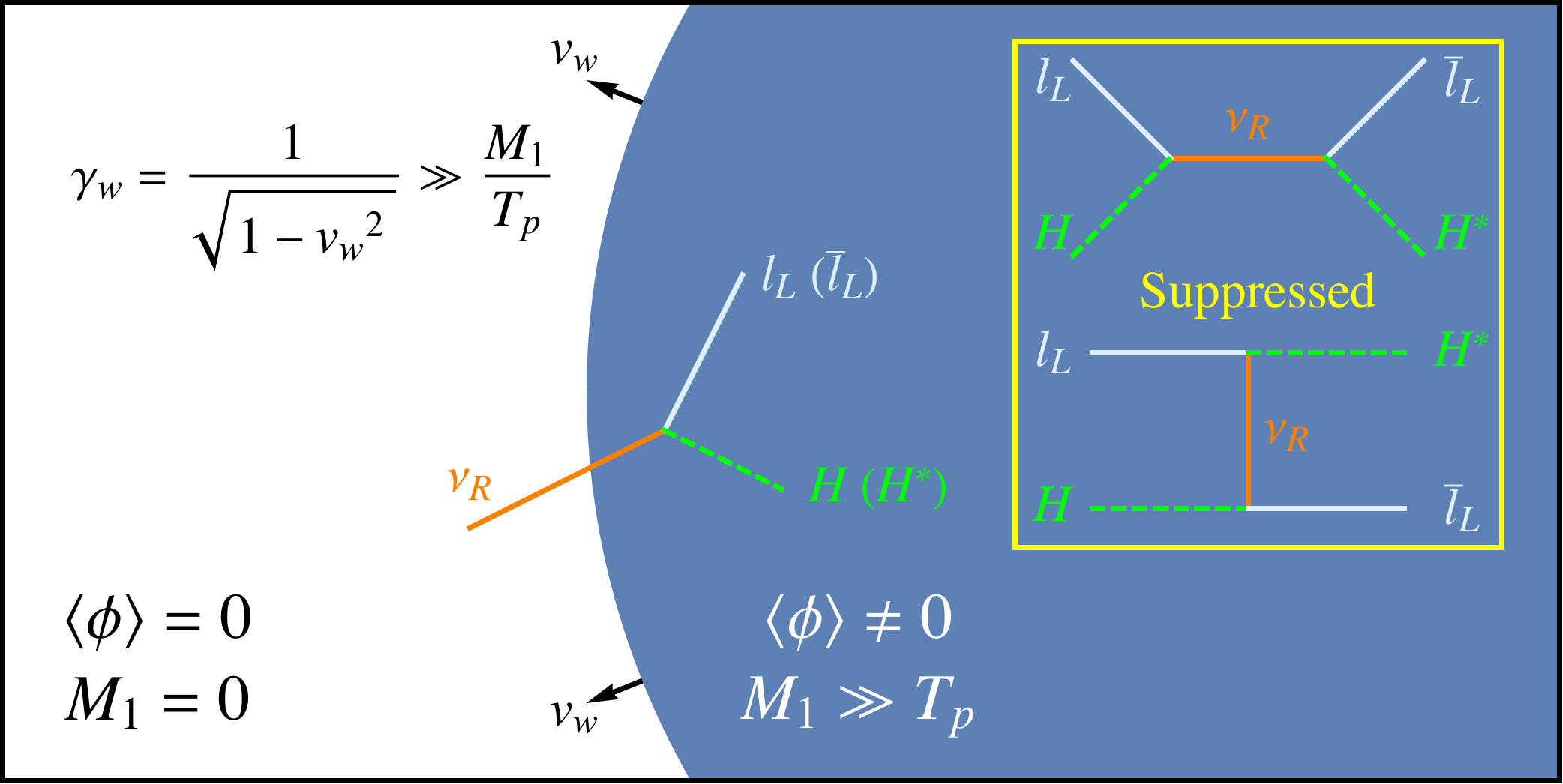}
\caption{The sketch of leptogenesis triggered by a FOPT. The blue and white regions represent the new vacuum bubble (in which $\ave{\phi}\neq0$) and the old vacuum background (in which $\ave{\phi}=0$), respectively. The FOPT occurs at temperature $T_p$, and the bubble expands at a wall velocity $v_w$. Inside a bubble, $\nu_R$ gains a huge mass $M_1\gg T_p$, such that the $\nu_R$'s that have penetrated the bubble decay quickly, generating the BAU. The possible washout effects (some of which are illustrated inside the yellow rectangle) are suppressed since $M_1/T_p\gg 1$.}
\label{fig:sketch}
\end{figure}

In this article, we propose a new scenario of leptogenesis, which is triggered by a first-order phase transition~(FOPT). The idea is quite simple: in many models such as the $B-L$ or Majoron model, RHNs obtain masses through the vacuum expectation value~(VEV) of a scalar field $\phi$. If in the early Universe $\phi$ experiences a FOPT from $\ave{\phi}=0$ to $\ave{\phi}\neq0$, then during the phase transition $\nu_R$ would be massless in the old vacuum, while massive in the new vacuum bubbles. If the mass gap is much larger than the FOPT temperature $T_p$, then the $\nu_R$'s that have penetrated into the new vacuum will be out of equilibrium and decay rapidly, generating the lepton asymmetry and hence the BAU. Since $M_1\gg T_p$, the washout effects are Boltzmann suppressed, and hence almost all the generated BAU can survive till today. The mechanism is sketched in Fig.~\ref{fig:sketch}. The idea of generating BAU via the fast decay of heavy particles crossing or being produced at the bubble wall is first proposed in Ref.~\cite{Baldes:2021vyz}, where the general features are discussed, and a benchmark model on a color triplet heavy scalar is given. Our work provides the first detailed study of applying this idea to leptogenesis. This scenario is distinct from the mechanisms involving the generation and diffusion of chiral and/or lepton asymmetry in the vicinity of bubble~\cite{Chung:2009cb,Chiang:2016vgf,Guo:2016ixx,DeVries:2018aul,Xie:2020wzn,Cline:2021dkf} (see also~\cite{Pascoli:2016gkf,Long:2017rdo}).\footnote{See Refs.~\cite{Pilaftsis:2008qt,Shuve:2017jgj} for leptogenesis during a second-order phase transition.}

One might concern that in case of $M_1/T_p\gg1$ the RHNs do not have sufficient energy to cross the wall, as the average kinetic energy of $\nu_R$ is $\mO(T_p)$. Were that true, most of the RHNs will be trapped in the old vacuum~\cite{Hong:2020est,Baker:2021nyl,Kawana:2021tde,Baker:2021sno,Arakawa:2021wgz}, and only a tiny fraction of them can be ``filtered'' to the new vacuum~\cite{Baker:2019ndr,Chway:2019kft,Chao:2020adk,Baker:2021zsf}, resulting in a much suppressed $\nu_R$ number density in the $\ave{\phi}\neq0$ phase, and the resultant BAU is also negligible, as pointed out in ~\cite{Shuve:2017jgj,Ahmadvand:2021vxs}. This issue, however, can be solved, provided that the bubbles are expanding in an ultra-relativistic velocity, i.e. $\gamma_w\equiv(1-v_w^2)^{-1/2}\gg1$. In that case, in the wall rest frame the RHNs have an average kinetic energy $\mO(\gamma_wT_p)$, and almost all the $\nu_R$'s can penetrate into the bubble, yielding an unsuppressed number density $\sim T_p^3$ inside the new vacuum. We will show that $\gamma_w\gg1$ can be easily achieved in a supercooling FOPT. See Refs.~\cite{Katz:2016adq,Azatov:2020ufh,Azatov:2021ifm,Azatov:2021irb} for other cosmological implications of the ultra-relativistic walls.

Compared with the conventional thermal leptogenesis scenario, our FOPT leptogenesis scenario has an enhanced RHN number density ($T_p^3$ instead of $(M_1T_p)^{3/2}e^{-M_1/T_p}$) and does not suffer from thermal washout effects (which in general suppress the BAU by a factor of $\mO(10^{-2})$). Therefore, naively we expect the CP violating phase needed by the FOPT scenario is much smaller than that in the conventional scenario, and hence the FOPT scenario is able to explain the BAU at a lower $M_1\lesssim 10^9$ GeV assuming the Davidson-Ibarra bound. However, the FOPT scenario suffers from the washout and dilution effects {\it after} the FOPT. This is because the ultra-relativistic wall requires a strong FOPT, which releases a large amount of latent heat and then reheats the Universe to a high temperature $T_{\rm rh}>T_p$. It is difficult to satisfy $M_1/T_{\rm rh}\gg1$, which is the condition to suppress the washout effect after reheating. In addition, the generated BAU will be diluted by a factor of $(T_p/T_{\rm rh})^3$.

In this article, we will provide a realization of the FOPT leptogenesis scenario in the {\it classically conformal} $B-L$ model with $M_1\gtrsim10^{11}$ GeV, taking account of the reheating washout and dilution effects. While in the same parameter space, the conventional thermal leptogenesis generates a BAU much smaller than the observed value. Therefore, our research extends the parameter space for leptogenesis. This article is organized as follows. Before moving to the concrete model building, we will first study the dynamics of the FOPT leptogenesis in Section~\ref{sec:dynamics}, keeping the discussions as general as we can. Then Section~\ref{sec:model_building} introduces a concrete extended $B-L$ model and demonstrates the parameter space realizing a FOPT leptogenesis scenario. The possible gravitational wave (GW) signals are also studied. Finally, we conclude in Section~\ref{sec:conclusion}.

\section{Dynamics of the FOPT leptogenesis}\label{sec:dynamics}

\subsection{Basic setup}

In this section, we do not specify a concrete model. The discussions apply to any model that contains the following two features. First, the RHNs $\nu_R^i$ have the Majorana Yukawa interaction
\be\label{massR}
\mL\supset-\sum_{i,j}\frac{1}{2}\left(\lambda_{R}^{ij}\bar\nu_R^{i,c}\nu_R^j\frac{\phi}{\sqrt{2}}+\hc\right),
\ee
where $i$, $j$ are family indices, $\phi$ is a real scalar field that experiences a FOPT from $\ave{\phi}=0$ to $\ave{\phi}=v_p$ at temperature $T_p$. Therefore, the RHNs are massless in the old vacuum but obtain masses $M_R^{ij}=\lambda_{R}^{ij}v_p/\sqrt{2}$ inside the new vacuum bubble. For simplicity we set $M_R^{ij}={\rm diag}\{M_1,M_2,M_3\}$ and let $\nu_R^1$ be the lightest RHN. The second feature is that the RHNs should couple to the SM leptons and bosons via the Dirac Yukawa interaction
\be\label{DiracYukawa}
\mL\supset-\sum_{i,j}\left(\lambda_D^{ij}\bar\ell_L^i\tilde H\nu_R^j+\hc\right),
\ee
where $\ell_L^i=(\nu_L^i,e_L^i)^T$ is the lepton doublet, and $\tilde H=i\tau^2H^*$ is the charge conjugation of the Higgs doublet. \Eq{DiracYukawa} allows the RHNs to decay via $\nu_R^i\to\ell_L^jH/\bar\ell_L^jH^*$, and hence to generate the lepton asymmetry.

The magnitude of the $\lambda_D^{ij}$ matrix can be estimated by the seesaw relation $m_\nu\approx|\lambda_D|^2v_{\rm EW}^2/(2M_R)$ as
\be\label{seesaw_lambdaD}
|\lambda_D|\approx10^{-2}\times\left(\frac{M_R}{10^{11}~{\rm GeV}}\right)^{1/2}\left(\frac{m_\nu}{0.05~{\rm eV}}\right)^{1/2},
\ee
where $v_{\rm EW}=246$ GeV is the Higgs VEV. The CP violating effect is characterized by the RHN decay width asymmetry
\be
\epsilon_i=\frac{\sum_j\Gamma(\nu_R^i\to\ell_L^j H)-\Gamma(\nu_R^i\to\bar\ell_L^j H^*)}{\sum_j\Gamma(\nu_R^i\to\ell_L^j H)+\Gamma(\nu_R^i\to\bar\ell_L^j H^*)},
\ee
which is related to the imaginary part of $(\lambda_D\lambda_D^\dagger)^2$. A nonzero $\epsilon_1$ is needed for the generation of BAU. According to the Davidson-Ibarra bound~\cite{Davidson:2002qv},
\be\label{DI_bound}
|\epsilon_1|\leqslant\frac{3}{8\pi}\frac{M_1(m_3-m_1)}{v_{\rm EW}^2}\approx10^{-5}\times\left(\frac{M_1}{10^{11}~{\rm GeV}}\right)\left(\frac{m_\nu}{0.05~{\rm eV}}\right).
\ee
We can see that $\epsilon_1$ is quite small even for a rather heavy $\nu_R^1$.

Above is the basic setup of the FOPT leptogenesis mechanism. When applying this mechanism, we allow a concrete model to have more ingredients, such as a $Z'$ boson from the gauged $U(1)_{B-L}$ group or other additional scalars and fermions. To realize our mechanism, three things must be checked. \underline{First}, right after penetration, the RHNs should decay instead of annihilating with each other, or scattering with the particles in the thermal bath. \underline{Second}, as the penetrated RHNs are typically boosted, so are the decay products, and it is necessary to check that they do not cause additional washout effects for the BAU. \underline{Third}, after the FOPT, the Universe is reheated to $T_{\rm rh}$ and we have to confirm the thermal bath washout effects are still Boltzmann suppressed even at this temperature. Also, the dilution factor $(T_p/T_{\rm rh})^3$ should be included. All those issues will be addressed one by one in the subsequent subsections.

\subsection{RHNs right after penetration}

In the vicinity of the bubble wall, we can model the bubble expansion as a one-dimension problem: the wall is a plane perpendicular to the $z$-axis and moving in a velocity $-v_w$ with $v_w>0$. The $z\to-\infty$ region is the old phase with $\ave{\phi}=0$, where the RHNs are assumed to be in thermal equilibrium. Therefore, the lightest RHN $\nu_R^1$ follows a boosted massless Fermi-Dirac distribution in the wall frame
\be
f_\s^{\rm wa}(p_x,p_y,p_z)=\frac{1}{e^{\gamma_w(E_0-v_wp_z)/T_p}+1},
\ee
where $E_0=\sqrt{p_x^2+p_y^2+p_z^2}$. The corresponding particle number density is
\be
n_\s^{\rm wa}=g_\nu\int\frac{d^3p}{(2\pi)^3}f_\s^{\rm wa}(p_x,p_y,p_z)=\gamma_w\times g_\nu\frac{3\zeta_3}{4\pi^2}T_p^3\equiv\gamma_w n_\s^{\rm pl},
\ee
where $\zeta_3\approx1.202$, and $g_\nu=2$ is the spin degeneracy factor. $n_\s^{\rm wa}$ is enhanced by a factor of $\gamma_w$ compared with $n_\s^{\rm pl}$ in the plasma frame, which can be understood as the Lorentz contraction of the volume element. Note that we use a superscript ``wa'' (``pl'') to label the wall frame (plasma frame), and a subscript ``$\s$'' (``$\h$'') to label the old vacuum with $\ave{\phi}=0$ (new vacuum with $\ave{\phi}=v_p$), respectively.

In the wall frame the average $z$-component momentum is $\ave{p_z}^{\rm wa}_{\s}=7\pi^4v_w\gamma_wT_p/(135\zeta_3)$, and hence the RHNs are boosted along the $+z$ direction. If $\gamma_w$ is large enough, $\ave{p_z}^{\rm wa}_{\s}$ is enhanced that most $\nu_R^1$'s have sufficient energy to overcome the mass gap $M_1$ between the new and old vacua. Hereafter we only consider the $\gamma_w\gtrsim M_1/T_p\gg1$ limit, then the $\nu_R^1$'s can generally cross the wall and enter the new vacuum. Due to energy conservation in the wall frame, after crossing the wall, in the new vacuum the average energy and momentum of $\nu_R^1$ should be
\be
\ave{E}^{\rm wa}_\h\sim\gamma_wT_p,\quad \ave{p_z}^{\rm wa}_\h\sim\sqrt{\gamma_w^2T_p^2-M_1^2}\sim\gamma_wT_p-\frac{M_1^2}{2\gamma_wT_p}.
\ee
Transforming back to the plasma frame, one obtains the typical $\nu_R^1$ energy and momentum after penetrating into the bubble
\be
\ave{E}^{\rm pl}_\h\sim M_1\frac{M_1}{T_p},\quad \ave{p_z}^{\rm pl}_\h\sim-M_1\frac{M_1}{T_p},
\ee
which means in the plasma frame the $\nu_R^1$'s that have entered the new vacuum are boosted in the $-z$ direction by a Lorentz factor of $\gamma_1\equiv M_1/T_p\gg1$. In other words, in the plasma frame, part of the wall kinetic energy is converted into the rest mass and kinetic energy of the RHNs that enter the bubble. This causes the energy loss of the wall and serves as a source of the friction force acting on the wall, as we will discuss in \Eq{P11}.

After entering the new vacuum, a $\nu_R^1$ may decay, or annihilate with another $\nu_R^1$, or scatter with the particles in the plasma. When calculating these interaction rates, it is convenient to work in the ``$\nu_R^1$ gas frame'', which is boosted along the $-z$ direction with a Lorentz factor $\gamma_1$. In that frame, the $\nu_R$'s are on average at rest, and with a relative velocity $v^{\rm ga}_{\rm rel}\sim T_p/M_1$ to each other, and the number density is $n^{\rm ga}_\h\approx\gamma_1n^{\rm pl}_\s$~\cite{Baldes:2021vyz}. In the gas frame, the $\nu_R^1$ decay rate is
\be\label{GammaD}
\Gamma_D=\frac{|\lambda_D|^2}{8\pi}M_1\approx\frac{m_\nu}{4\pi}\left(\frac{M_1}{v_{\rm EW}}\right)^2,
\ee
where we have assumed one-flavor SM final state for simplicity, and the second approximate equality is from the seesaw relation.

Depending on the concrete model, the $\nu_R^1$'s can annihilate with each other via various channels. For example, the Majorana interaction \Eq{massR} induces the annihilation to a pair of scalars, i.e. $\nu_R^1\nu_R^1\to\phi\phi$, if kinematically allowed, while the Dirac interaction \Eq{DiracYukawa} induces $\nu_R^1\nu_R^1\to\ell_L\bar\ell_L/HH^*$. If the model is embedded into a gauged $U(1)_{B-L}$ group, then the $\nu_R^1\nu_R^1\to Z'^{*}\to f\bar f$ and $\nu_R^1\nu_R^1\to Z'Z'$ channels may be important, where $Z'$ and $f$ denote the $U(1)_{B-L}$ gauge boson and SM fermions, respectively.  Then the annihilation rate can be expressed as
\be
\Gamma_{\rm ann}=\sum_Xn_\h^{\rm ga}\ave{\sigma_{\nu_R^1\nu_R^1\to X}v_{\rm rel}}_{\rm ga},
\ee
summing over all possible annihilation final states. The subscript ``ga'' of $\ave{\sigma v}$ is to remind us that this is an average performed in the gas frame. $\Gamma_{\rm ann}$ scales as $T_p^3/M_1^2$.

Another possible fate of the penetrated $\nu_R^1$'s is to scatter with the particles in the plasma. The Dirac Yukawa interaction can mediate scattering channels such as $\nu_R^1\ell_L\to q_L\bar t_R$ or $\nu_R^1t_R\to q_L\bar\ell_L$ and their charge conjugations and crossings. The corresponding interaction rates are
\be
\Gamma_{\rm sca}=\sum_{a,X}\gamma_1n_a^{\rm pl}\ave{\sigma_{\nu_R^1a\to X}}_{\rm ga},
\ee
summing over all possible initial states $a$ and final states $X$. In the gas frame, the plasma species $a$ is boosted by a Lorentz factor of $\gamma_1$, therefore the number density is enhanced by $\gamma_1$ compared with $n_a^{\rm pl}\sim T_p^3$ in the plasma frame, and we have taken the relative velocity between $\nu_R^1$ and $a$ to be approximately 1. The scattering cross section $\langle\sigma_{\nu_R^1a\to X}\rangle_{\rm ga}$ scales as $1/M_1^2$, thus $\Gamma_{\rm sca}\sim T_p^2/M_1$.

For the sake of leptogenesis, we want the $\nu_R^1$'s to decay rather than annihilate with each other or scatter with the particles in the plasma, i.e.
\be\label{decay}
\Gamma_D>\Gamma_{\rm ann},\quad \Gamma_D>\Gamma_{\rm sca}.
\ee
Under this condition, the $\nu_R^1$'s swept by the bubble wall decay immediately and generate a BAU of
\be\label{YBp}
Y_B^p=-c_s\epsilon_1\frac{n_\s^{\rm pl}}{s}=-c_s\epsilon_1\frac{135\zeta_3}{4\pi^4g_*},
\ee
where $s=(2\pi^2/45)g_*T^3$ is the entropy density with $g_*\approx100$ the number of relativistic degrees of freedom, and $c_s=28/79$ is the conversion factor from the lepton asymmetry to the BAU. As the upper limit of CP asymmetry $\epsilon_1$ is constrained by \Eq{DI_bound}, we see that the maximal value of BAU is proportional to $M_1$.

\subsection{The boosted decay products of RHNs}

In the plasma frame, the $\nu_R^1$'s in new vacuum are moving along the $-z$ direction with a typical energy $E_1=\gamma_1M_1=M_1^2/T_p$. The decay products $\ell_LH/\bar\ell_LH^*$ share the same order of energy and hence are also boosted. These out-of-equilibrium SM particles interact with other SM particles in the plasma, causing cascade scatterings, which might reduce the BAU. Following Ref.~\cite{Baldes:2021vyz}, we model the energy of the particles that in the $n$-th step cascade scattering as $E_1/2^n$. The washout effect is mainly from the possibility that the energetic particles fuse to an on-shell RHN, i.e. $\ell_LH\to\nu_R^1\to\bar\ell_LH^*$, and the corresponding rate can be estimated as~\cite{Baldes:2021vyz}
\be
\Gamma_{\rm on}\approx\frac{\Gamma_{\ell_L H}\Gamma_{\bar\ell_L H^*}}{\Gamma_D}\frac{M_1T_p}{E_1^2}\exp\left\{-\frac{M_1^2}{4E_1T_p}\right\}\approx\frac{2^{2n}T_p^3}{4M_1^3}\Gamma_De^{-2^n/4},
\ee
where we have approximated $\Gamma_{\ell_LH}\approx\Gamma_{\bar\ell_LH^*}\approx\Gamma_D/2$. We can see that the washout rate decreases very quickly as $n$ increases, so we only need to account for the first step of scattering, i.e. $n=1$.

Being charged under the SM gauge groups, the boosted $\ell_L/\bar\ell_L$ and $H/H^*$ particles also thermalize via the elastic EW scattering with the SM particles in the plasma. The thermalization rate can be estimated by calculating the energy loss of a boosted lepton in an elastic scattering with another SM particle in the thermal bath. The two incoming particles have momenta
\be
p_1^\mu=\left(\frac{E_1}{2^n},0,0,\frac{E_1}{2^n}\right),\quad p_2^\mu=\left(T_p,0,0,-T_p\right),
\ee
respectively, and they scatter through exchanging a $t$-channel $W/Z$ boson. It is straightforward to show that the energy loss of the boosted lepton is $\delta E_1\approx-\hat t/(4T_p)$ in the plasma frame, and the scattering cross section is
\be
\frac{d\sigma}{d\hat t}=\frac{1}{16\pi\hat s^2}|i\mM|^2\approx\frac{1}{16\pi\hat s^2}\frac{g_2^4\hat s^2}{\hat t^2}=\frac{\pi\alpha_W^2}{\hat t^2},
\ee
where $g_2$ is the gauge coupling of the $SU(2)_L$ group, and $\hat s$, $\hat t$ are the Mandelstam variables. Therefore, the thermalization rate can be estimated as~\cite{Baldes:2021vyz}
\be
\Gamma_{\rm th}=\frac{n_{\rm EW}^{\rm pl}}{E_1/2^n}\int_{-\hat s}^{-m_W^2}d\hat t\frac{d\sigma}{d\hat t}\delta E_1=\frac{\zeta_3g_{\rm EW}2^n\alpha_W^2T_p^3}{4\pi M_1^2}\ln\frac{3M_1^2}{5\pi 2^n\alpha_WT_p^2},
\ee
where $n_{\rm EW}^{\rm pl}$ is the number density of the particles that participate in such EW elastic scattering, and the corresponding number of degrees of freedom is $g_{\rm EW}=46$ including the SM fermions and gauge bosons as well as the Higgs doublet. The upper limit of integration of $\hat t$ is set to $-m_W^2$ to avoid infrared divergence, where $m_W^2=20\pi\alpha_WT_p^2/3$ is the thermal mass of the $W$ boson~\cite{Weldon:1982aq}. We see that $\Gamma_{\rm th}$ increases rapidly with $n$.

To avoid washout from the boosted decay products, we require
\be\label{boosted_decay}
\Gamma_{\rm th}\big|_{n=1}>\Gamma_{\rm on}\big|_{n=1},\quad \Gamma_{\rm th}\big|_{n=1}>H_p,
\ee
where $H_p$ is the Hubble constant at the FOPT temperature. Note that $H_p$ is not solely determined by temperature, as the vacuum energy from the potential could dominate the energy of the Universe in the case of a supercooling FOPT. Once these inequalities are satisfied, the boosted decay products $\ell_L H/\bar\ell_L H^*$ thermalize very quickly, and the washout effect is completely negligible.

\subsection{Reheating after the FOPT completes}

The latent heat released from a FOPT will reheat the Universe to a new temperature $T_{\rm rh}=(1+\alpha)^{1/4}T_p$, where $\alpha$ is the ratio of latent heat to the radiation energy density of the Universe, whose detailed definition will be given in Section~\ref{sec:foptleptogenesis}. Since our scenario needs a strong FOPT to provide fast moving bubble walls, typically $\alpha\gg1$, the reheating temperature could be very high, such that the $B-L$ violating interactions are active again, erasing the generated $B-L$ asymmetry as the situation in the conventional thermal leptogenesis.

The first type of dangerous processes is the thermally produced RHNs. For the inverse decay, i.e. $\ell_L H/\bar\ell_LH^*\to \nu_R^i$, the simplified Boltzmann equation gives
\be
\frac{dY_{\nu_R^i}}{dt}=\frac1s\int\frac{d^3p}{(2\pi)^32E_i}e^{-E_i/T}M_i\Gamma_{D,i}=\frac{\Gamma_{D,i}M_i^2T}{4\pi^2s}K_1\left(\frac{M_i}{T}\right),
\ee
where $Y_{\nu_R^i}=n_{\nu_R^i}/s$ is the yield of the $i$-th generation of RHN, $E_i\equiv\sqrt{|\p|^2+M_i^2}$ is the on-shell energy, $\Gamma_{D,i}$ is the decay width of $\nu_R^i$, and $K_1$ is the modified Bessel function of the first kind. By this, the inverse decay rate can be estimated as
\be\label{GammaID}
\Gamma_{\rm ID}^i=\frac{\Gamma_{D,i}M_i^2T_{\rm rh}}{4\pi^2s_{\rm rh}}K_1\left(\frac{M_i}{T_{\rm rh}}\right),
\ee
where $s_{\rm rh}=s|_{T_{\rm rh}}$. We should have $\Gamma_{\rm ID}^i<H_{\rm rh}$ such that the RHNs are not thermally produced after the FOPT, where the Hubble constant $H_{\rm rh}=2\pi\sqrt{\pi g_*/45}T_{\rm rh}^2/M_{\rm Pl}$ with $M_{\rm Pl}=1.22\times10^{19}$ GeV, as the Universe is in a radiation era after the FOPT. The rate of RHNs being produced in pair in the plasma can be estimated as
\be
\Gamma_{\rm pr}^i=\sum_Xn_{\nu_R^i}^{\rm eq}\ave{\sigma_{\nu_R^i\nu_R^i\to X}}=\sum_X2\frac{M_i^2T_{\rm rh}}{2\pi^2}K_2\left(\frac{M_i}{T_{\rm rh}}\right)\ave{\sigma_{\nu_R^i\nu_R^i\to X}},
\ee
where $n_{\nu_R^i}^{\rm eq}$ is the equilibrium distribution of $\nu_R^i$ in the plasma whose concrete expression is given in the second equality with $K_2$ being the modified Bessel function of the second kind. Depending on the model, the pair production channels could include $\nu_R^i\nu_R^i\to Z'^*\to f\bar f$, $\nu_R^i\nu_R^i\to Z'Z'$, $\nu_R^i\nu_R^i\to Z'\phi$, $\nu_R^i\nu_R^i\to \phi\phi$, etc.

We require
\be\label{thermal_washout}
\Gamma_{\rm ID}^i<H_{\rm rh},\quad \Gamma_{\rm pr}^i<H_{\rm rh},
\ee
to avoid thermal bath washout after the FOPT reheating. Both $\Gamma_{\rm ID}^i$ and $\Gamma_{\rm pr}^i$ are suppressed by the Bessel functions, which are $K_j(z)\sim e^{-z}\sqrt{\pi/(2z)}$ for $z\gg1$. Therefore, $M_i/T_{\rm rh}\gg1$ could exponentially suppress those washout effects. In other words, we need $T_{\rm rh}=(1+\alpha)^{1/4}T_p$ still small compared with the RHN masses; this is, however, in tension with the requirement of a strong supercooling FOPT which generally leads to $\alpha\gg1$.

Provided that Eqs.~(\ref{decay}), (\ref{boosted_decay}) and (\ref{thermal_washout}) are satisfied, the FOPT leptogenesis scenario is realized. Namely, the $\nu_R^1$'s that have entered the new vacuum bubble during the FOPT will decay and generate the lepton asymmetry, which is not washed out by the plasma. The BAU survives today would be
\be\label{YB_final}
Y_B=-c_s\epsilon_1\frac{135\zeta_3}{4\pi^4g_*}\left(\frac{T_p}{T_{\rm rh}}\right)^3,
\ee
which is diluted by a factor of $(T_p/T_{\rm rh})^3$ compared to \Eq{YBp}, due to the entropy production of the FOPT reheating. For a successful FOPT leptogenesis, $Y_B$ should reach the observed BAU, i.e. $Y_B^{\rm obs}\approx0.9\times10^{-10}$~\cite{ParticleDataGroup:2020ssz}.

In summary, in the FOPT leptogenesis scenario, the FOPT should be strong to provide fast expanding bubbles, which sweep the RHN into the new vacuum. Therefore, the abundant massless $\nu_R^1$ density in the old vacuum can be directly transferred into the new vacuum, where the $\nu_R^1$'s are so massive that their out-of-equilibrium decay can generate the BAU without the washout effects. However, the reheating effects from the strong FOPT might cause additional washout and dilution effects, and hence the application of this mechanism requires a highly non-trivial tradeoff between strong FOPT and reheating. A concrete model that succeeds to realize the FOPT leptogenesis scenario is given in the next section.

\section{An extended classically conformal $B-L$ model}\label{sec:model_building}

\subsection{The model and particle spectrum}

The conventional (or say, minimal) $B-L$ model~\cite{Davidson:1978pm,Marshak:1979fm,Mohapatra:1980qe,Davidson:1987mh} is defined by gauging the $U(1)_{B-L}$ group and introducing three generation of RHNs (with $B-L$ quantum number $X=-1$) for gauge anomaly cancellation, and one complex scalar field $\Phi=(\phi+i\eta)/\sqrt2$ charged as $X=2$ to break the $U(1)_{B-L}$ spontaneously. In this work, we extend the model with one more complex scalar $S$ which has the same quantum number with $\Phi$. The relevant Lagrangian can be written as
\be\label{LB-L}\begin{split}
\mL_{B-L}=&~\sum_{i}\bar\nu_R^ii\slashed{D}\nu_R^i-\frac12\sum_{i,j}\left(\lambda_R^{ij}\bar\nu_R^{i,c}\Phi\nu_R^j+\hc\right)
-\sum_{i,j}\left(\lambda_D^{ij}\bar\ell_L^i\tilde H\nu_R^j+\hc\right)\\
&~+D_\mu\Phi^\dagger D^\mu\Phi+D_\mu S^\dagger D^\mu S-V(\Phi,S)-\frac14Z'_{\mu\nu}Z'^{\mu\nu},
\end{split}\ee
where $D_\mu=\partial_\mu-ig_{B-L}XZ_\mu'$ is the $U(1)_{B-L}$ gauge covariant derivative. For simplicity, we take $\lambda_R^{ij}={\rm diag}\{\lambda_{R,1},\lambda_{R,2},\lambda_{R,3}\}$. Note that the SM fermions are also charged under the $U(1)_{B-L}$ group, with the quarks having $X=1/3$ and the leptons having $X=-1$. The reason why we have to extend the minimal $B-L$ model will be given in Section~\ref{sec:foptleptogenesis}. In principle, $S$ can also couple to RHNs via $\bar\nu_R^{i,c}S\nu_R^j$; however, as we will see, $S$ never gets a VEV, thus it does not contribute to the RHN mass. On the other hand, $S$ can provide extra CP violating phase to $N_1$ decay~\cite{LeDall:2014too,Alanne:2018brf}. We do not consider such CP asymmetry enhancement effects here, as they are irrelevant to the core of our FOPT leptogenesis mechanism.

As for the scalar potential $V(\Phi,S)$, we adopt the classically conformal assumption~\cite{Iso:2009ss,Iso:2009nw,Das:2015nwk} as it is known that this kind of potential favors a strong supercooling FOPT~\cite{Jinno:2016knw,Iso:2017uuu,Marzo:2018nov,Ellis:2019oqb,Bian:2019szo,Ellis:2020nnr,Jung:2021vap}. At three level, the potential is
\be
V_{\rm tree}(\Phi,S)=\lambda_\phi|\Phi|^4+\lambda_s|S|^4+\lambda_{\phi s}|\Phi|^2|S|^2,
\ee
where only dimensionless quartic couplings are involved. The one-loop contributions from $Z'$ and $\nu_R^i$~\cite{Iso:2009ss,Iso:2009nw,Das:2015nwk} and $S$~\cite{Haruna:2019zeu,Hamada:2020wjh,Hamada:2021jls,Kawana:2022fum} induce a Colman-Weinberg potential for $\Phi$, which in the unitary gauge can be written as
\be\label{CC_B-L_CWN_S}
V(\phi)=V_0+\frac{B}{4}\phi^4\left(\ln\frac{\phi}{v_\phi}-\frac14\right),
\ee
where
\be\label{B}
B=\frac{6}{\pi^2}\left(\frac{\lambda_{\phi s}^2}{96}+g_{B-L}^4-\sum_i\frac{\lambda_{R,i}^4}{96}\right),
\ee
is a positive constant. This potential has the a minimum at $\ave{\phi}=v_\phi\neq0$, which breaks the $U(1)_{B-L}$ symmetry spontaneously and provides masses for the particles in \Eq{LB-L} as follows
\be
M_{Z'}=2g_{B-L}v_\phi,\quad M_i=\lambda_{R,i}\frac{v_\phi}{\sqrt2},\quad M_\phi=\sqrt{B}v_\phi,\quad M_S=\frac{1}{\sqrt2}\sqrt{\lambda_{\phi s}}v_\phi.
\ee
The vacuum energy is adopted as $V_0=Bv_\phi^4/16$ to have $V(v_\phi)=0$.

\subsection{FOPT and ultra-relativistic bubble walls}

At finite temperature, the potential receives corrections from the one-loop thermal integrals and daisy resummation terms 
\begin{multline}
\Delta V_T(\phi,T)=2\frac{T^4}{2\pi^2}J_B\left(\frac{\lambda_{\phi s}\phi^2}{2T^2}\right)+3\frac{T^4}{2\pi^2}J_B\left(\frac{4g_{B-L}^2\phi^2}{T^2}\right)+2\sum_i\frac{T^4}{2\pi^2}J_F\left(\frac{\lambda_{R,i}^2\phi^2}{2T^2}\right)\\
-2\frac{T}{12\pi}\frac{\lambda_{\phi s}^{3/2}}{2^{3/2}}\left[\left(\phi^2+\frac{T^2}{12}\right)^{3/2}-\phi^3\right]
-\frac{2g_{B-L}^3}{3\pi}T\left[\left(\phi^2+T^2\right)^{3/2}-\phi^3\right],
\end{multline}
where thermal integral functions are defined as
\be
J_{B/F}(y)=\pm\int_0^\infty dxx^2\ln\left(1\mp e^{-\sqrt{x^2+y}}\right).
\ee
Therefore, the complete one-loop thermal potential is
\be\label{VT}
V_T(\phi,T)=V(\phi)+\Delta V_T(\phi,T),
\ee
which can trigger a FOPT in the early Universe. The FOPT of the minimal classically conformal $B-L$ model has already been extensively studied~\cite{Jinno:2016knw,Iso:2017uuu,Marzo:2018nov,Ellis:2019oqb,Bian:2019szo,Ellis:2020nnr}, and in this work we use homemade codes to derive the FOPT dynamics of our extended $B-L$ model. 

When $T$ is sufficiently high, the Universe stays in the $U(1)_{B-L}$ preserving vacuum $\phi=0$. At the critical temperature $T_c$, the potential $V_T(\phi,T)$ develops another degenerate vacuum in $\phi=v_c$. When $T$ falls below $T_c$, the $U(1)_{B-L}$ breaking vacuum is energetically preferred, i.e. $\ave{\phi}=v(T)$ becomes the true vacuum and we have $v(T_c)=v_c$ and $v(0)=v_\phi$. The Universe then acquires a decay probability~\cite{Linde:1981zj}
\be
\Gamma(T)\sim T^4\left(\frac{S_3(T)}{2\pi T}\right)^{3/2}e^{-S_3(T)/T},
\ee
to the true vacuum, where $S_3(T)$ is the action of the bounce solution, which we numerically resolve from $V_T(\phi,T)$ based on the shooting algorithm. When the decay probability in a Hubble volume and a Hubble time scale reaches $\mO(1)$, new vacuum bubbles start to nucleate. Given $\Gamma(T)$, the volume fraction of the old vacuum in the Universe is~\cite{Guth:1979bh,Guth:1981uk}
\be\label{IT}
p(T)\equiv e^{-I(T)}=\exp\left\{-\frac{4\pi}{3}\int_{T}^{T_c}dT'\frac{\Gamma(T')}{T'^4H(T')}\left[\int_{T}^{T'}d\tilde T\frac{1}{H(\tilde T)}\right]^3\right\},
\ee
where we have taken the bubble velocity $v_w\to1$, and the Hubble constant is given by the Friedmann equation
\be\label{supercooling_H_sim}
H^2(T)=\frac{8\pi}{3M_{\rm Pl}^2}\left(\frac{\pi^2}{30}g_* T^4+V_0\right).
\ee
By definition $p(T_c)=1$. When $T$ decreases, $p(T)\to0$, and the Universe transfers entirely to the new vacuum, completing the FOPT.

The milestone that the new vacuum bubbles form an infinite connected cluster is called {\it percolation}, and it happens at $p(T_p)=0.71$~\cite{rintoul1997precise}, which defines the percolation temperature $T_p$ and VEV $v_p=v(T_p)$. We will calculate the leptogenesis at this temperature.\footnote{Note that the large vacuum energy released from a supercooled FOPT can lead to a short vacuum domination era. We have checked that the FOPT can complete via verifying~\cite{Ellis:2018mja}
\be
H(T)\left(3+T\frac{dI(T)}{dT}\right)\Big|_{T_p}<0,
\ee
where $I(T)$ is defined in \Eq{IT}.} Define
\be
\Delta V(T)=V_T(0,T)-V_T(v(T),T),
\ee
as the positive free energy difference between the true and false vacuum, and let $\Delta V_p\equiv \Delta V(T_p)$. The behavior of wall velocity is determined by vacuum pressure $\Delta V_p$ and the leading-order (LO) friction~\cite{Bodeker:2009qy}
\be\label{P11}
\mP_{1\to1}=\left(\lambda_{\phi s}+12g_{B-L}^2+\frac{1}{2}\sum_i\lambda_{R,i}^2\right)\frac{v_p^2T_p^2}{24},
\ee
which comes from the mass differences of $S$, $Z'$ and RHNs between the two sides of the bubble wall. If
\be
\Delta V_p>\mP_{1\to1},
\ee
then the wall will be accelerated up to a high velocity that is very close to the speed of light, providing necessary condition for our FOPT leptogenesis scenario.

When $\gamma_w\gg1$, the beyond LO contributions to the friction force become important. Ref.~\cite{Bodeker:2017cim} performs the first next-to-leading-order (NLO) calculation, showing that the emission of gauge bosons when particles cross the wall can induce a friction force scaling as $\mP_{1\to2}\propto\gamma_w$, preventing the bubble walls from runaway. Recently, friction force on the wall is studied in many literatures~\cite{Hoche:2020ysm,Gouttenoire:2021kjv,Ai:2021kak,Cai:2020djd,Wang:2022txy,Laurent:2022jrs}, and we take the results of Refs.~\cite{Hoche:2020ysm,Gouttenoire:2021kjv} to calculate the evolution of bubble walls. While both two works consider the resummation effect of the $1\to N$ emission of gauge bosons, they obtain different friction pressures $\mP_{1\to N}$. Ref.~\cite{Hoche:2020ysm} shows $\mP_{1\to N}\propto\gamma_w^2$, however, Ref.~\cite{Gouttenoire:2021kjv} gives $\mP_{1\to N}\propto\gamma_w$. More concretely, applying to our model we find
\be\label{p1n_2}
\mP_{1\to N}^{\text{\scriptsize\cite{Hoche:2020ysm}}}\approx\gamma_w^2\left[\left(4\times2\times3\times\frac19+4+2\right)\times3\right]\frac{3\zeta_3(2\ln2-1)}{32\pi^4}g_{B-L}^2T_p^4,
\ee
and
\be\label{p1n_1}
\mP_{1\to N}^{\text{\scriptsize\cite{Gouttenoire:2021kjv}}}\approx\gamma_w\left[\left(4\times2\times3\times\frac19+4+2\right)\times3\right]\frac{3\kappa\zeta_3}{8\pi^4}g_{B-L}^3v_p T_p^3\ln\frac{v_p}{T_p},
\ee
where only the dominant SM fermion contributions are included, and $\kappa\approx4$.

The wall stops accelerating when the friction force balances the vacuum pressure, i.e. $\Delta V_p=\mP_{1\to 1}+\mP_{1\to N}$. Therefore, given the resummed friction force $\mP_{1\to N}$, we are able to derive the terminal wall velocity
\be
\gamma_{\rm eq}^{\text{\scriptsize\cite{Hoche:2020ysm}}}=\sqrt{\frac{\Delta V_p-\mP_{1\to1}}{\mP_{1\to N}^{\text{\scriptsize\cite{Hoche:2020ysm}}}/\gamma_w^2}};\quad \gamma_{\rm eq}^{\text{\scriptsize\cite{Gouttenoire:2021kjv}}}=\frac{\Delta V_p-\mP_{1\to1}}{\mP_{1\to N}^{\text{\scriptsize\cite{Gouttenoire:2021kjv}}}/\gamma_w}.
\ee
However, the wall might have not yet reached the terminal velocity at $T_p$. We use the method from Refs.~\cite{Ellis:2019oqb,Ellis:2020nnr} to evaluate the evolution of the wall velocity and confirm that for the parameter space of interest $\gamma_w\big|_{T_p}\gg M_1/T_p$ is indeed satisfied for either choice of $\mP_{1\to N}$, and hence all the $\nu_R^1$'s can penetrate the wall, which is the necessary condition for the FOPT leptogenesis.

\subsection{FOPT leptogenesis and the GW signals}\label{sec:foptleptogenesis}

Given the FOPT environment with ultra-relativistic bubble walls that can sweep all the RHNs into the new vacuum, we apply Eqs.~(\ref{decay}), (\ref{boosted_decay}) and (\ref{thermal_washout}) in Section~\ref{sec:dynamics} to our model \Eq{LB-L} to ensure that the penetrated $\nu_R^1$'s indeed decay before annihilating and scattering, and the boosted decay products thermalize instead of erasing the $B-L$ asymmetry, and the reheating temperature is still significantly below the RHN masses so that the thermal washout processes are Boltzmann suppressed.

As for the $\nu_R^1$'s right after penetration, the possible annihilation channels include $\nu_R^1\nu_R^1\to\ell_L\bar\ell_L/HH^*$, $\nu_R^1\nu_R^1\to Z'^{*}\to f\bar f$ with $f$ being the SM fermions, and $\nu_R^1\nu_R^1\to Z'Z'/Z'\phi/\phi\phi$. We calculate the corresponding annihilation rates using the {\tt FeynCalc} package~\cite{Mertig:1990an,Shtabovenko:2016sxi,Shtabovenko:2020gxv} and check that they are consistent with those in Refs.~\cite{Blanchet:2009bu,Heeck:2016oda,Duerr:2016tmh}. The $\nu_R^1$ scattering processes include $\nu_R^1\ell_L\to q_L\bar t_R$, $\nu_R^1t_R\to q_L\bar\ell_L$ and their charge conjugations and crossing diagrams. $\Gamma_D>\Gamma_{\rm ann}$ and $\Gamma_D>\Gamma_{\rm sca}$ are required for the fast decay of $\nu_R^1$, see \Eq{decay}. After decay, $\Gamma_{\rm th}>H_p$ and $\Gamma_{\rm th}>\Gamma_{\rm on}$ are needed for the decay products to thermalize quickly and do not reduce the generated $B-L$ asymmetry, see \Eq{boosted_decay}. The reheating temperature is $T_{\rm rh}=(1+\alpha)^{1/4}T_p$, where
\be\label{alphabeta}
\alpha=\frac{1}{g_*\pi^2T_p^4/30}\left(\Delta V(T)-\frac{T}{4}\frac{\partial \Delta V(T)}{\partial T}\right)\Big|_{T_p},
\ee
is the ratio of the FOPT latent heat to the radiation energy density. After confirming that the washout effects are suppressed even after reheating, i.e. \Eq{thermal_washout}, the eventual generated BAU is given by \Eq{YB_final}.

As stated in the Introduction, it is challenging to strike a balance between a strong FOPT and a not-so-strong reheating. A supercooling FOPT can provide fast-moving bubble walls, but the resultant large latent heat will push the Universe to a high $T_{\rm rh}$ that the thermal washout processes are active again, reducing any $B-L$ asymmetry generated during the FOPT. Especially, we find that it is in general difficult for the minimal classically conformal $B-L$ model~\cite{Iso:2009ss,Iso:2009nw} to realize the mechanism. In that model, the coefficient $B$ in \Eq{B} is determined only by $g_{B-L}$ and $\lambda_{R,i}$ that
\be\label{simB}
B\xrightarrow[B-L]{\rm Minimal}\frac{6}{\pi^2}\left(g_{B-L}^4-\sum_i\frac{\lambda_{R,i}^4}{96}\right)=\frac{3}{8\pi^2v_\phi^4}\left(M_{Z'}^4-\sum_i\frac{2M_{i}^4}{3}\right).
\ee
Therefore, $M_{Z'}\gtrsim M_i$ is required for a positive $B$ to ensure the vacuum stability. In addition, the FOPT requires a sizable $B$ to generate the potential barrier, and this implies a sizable $g_{B-L}$ and hence $M_{Z'}$ dominates \Eq{simB}. On the other hand, for a supercooling FOPT, the cosmic energy density is dominated by the vacuum energy and hence $T_{\rm rh}\sim V_0^{1/4}\sim B^{1/4}v_\phi\sim g_{B-L}v_\phi\sim M_{Z'}$. To have $M_1/T_{\rm rh}\gg1$ after reheating, we must have $M_1/M_{Z'}\gg1$, which is in contrast with the vacuum stability and FOPT conditions. We confirm this qualitative argument by a detailed numerical scan. Therefore, we extend the model with one extra scalar $S$, as we did in \Eq{LB-L}. In this new model, the contribution to $B$ can be dominated by the scalar portal coupling $\lambda_{\phi s}$, and the reheating temperature is no longer directly related to $M_{Z'}$.

\begin{figure}
\centering
\includegraphics[scale=0.45]{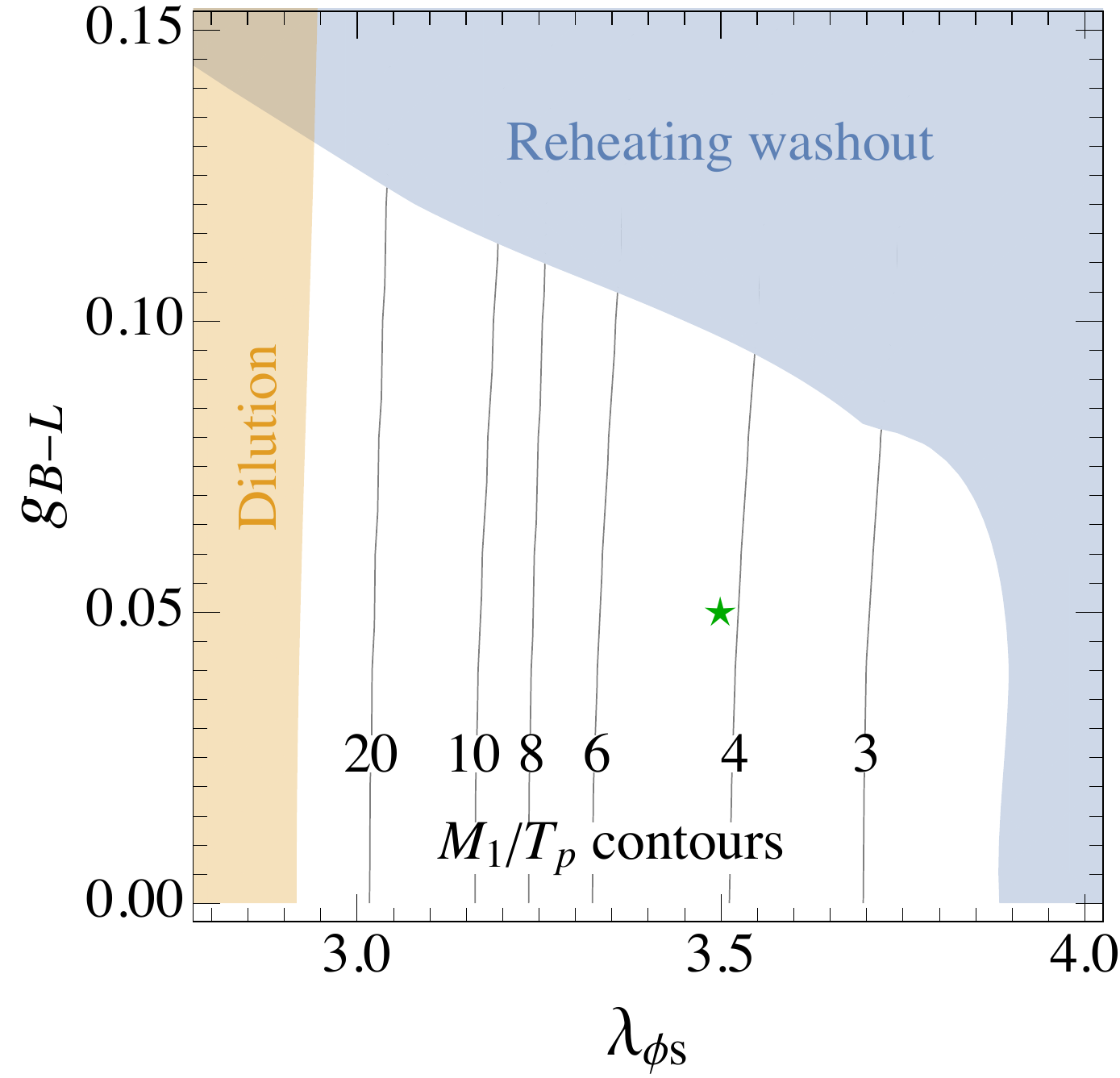}\qquad
\includegraphics[scale=0.45]{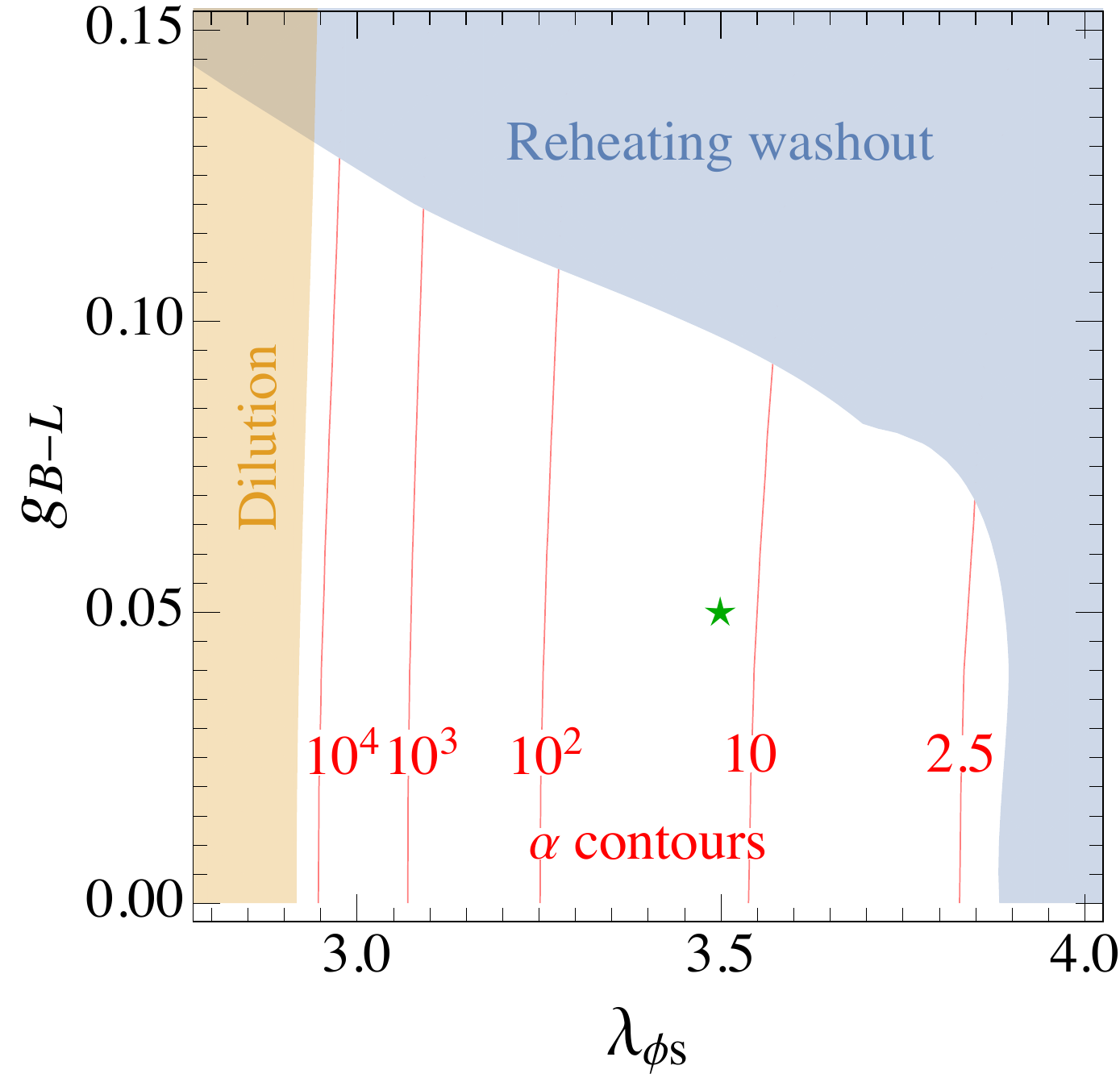}
\caption{The allowed parameter space of the FOPT leptogenesis scenario is shown in white region, for $M_1=2.5\times10^{11}$ GeV, $\lambda_{R,1}=0.3$ and $\lambda_{R,2}=\lambda_{R,3}=4\lambda_{R,1}$. The blue and orange shaded regions are excluded by thermal washout and dilution effects after the FOPT reheating, respectively. The $M_1/T_p$ and $\alpha$ contours are shown in the left and right panels, respectively. The green star is the benchmark adopted for GW calculation, see Fig.~\ref{fig:gws} for details.}
\label{fig:leptogenesis}
\end{figure}

For our extended $B-L$ model, we start from $M_1=10^9$ GeV and gradually increase it to seek for viable parameter space for the FOPT leptogenesis. The most stringent constraints for the scenario come from the washout effects after reheating, especially $\nu_R^i\nu_R^i\to Z'\phi$ and $\nu_R^i\nu_R^i\to Z'^{*}\to f\bar f$. Even in case that the reheating washout effects are suppressed, the BAU is usually diluted by the large $\alpha$ to be lower than the experimentally observed value. Therefore, we have to increase $M_1$ to $M_1\gtrsim10^{11}$ GeV to generate a large BAU. An example is shown in Fig.~\ref{fig:leptogenesis} with
\be
M_1=2.5\times10^{11}~{\rm GeV}, \quad \lambda_{R,1}=0.3,\quad \lambda_{R,2}=\lambda_{R,3}=4\lambda_{R,1},
\ee
fixed, and scanning over $\lambda_{\phi s}$ and $g_{B-L}$. The parameter space with successful FOPT leptogenesis, i.e. can provide $Y_B\geqslant Y_B^{\rm obs}$ for the $\epsilon_1$ within the Davidson-Ibarra bound, is plotted as the white region covered by the $M_1/T_p$ (left panel) and $\alpha$ (right panel) contours. We see $\alpha\gg1$ for most of the parameter space, implying a strong FOPT with vacuum energy dominance. The blue shaded region cannot realize FOPT leptogenesis because the thermal washout processes are active after reheating, where the $g_{B-L}\gtrsim0.1$ region is ruled out by the $\nu_R^i\nu_R^i\to Z'^*\to f\bar f$ annihilation, while the $\lambda_{\phi s}\gtrsim3.9$ region is excluded by the $\nu_R^i\nu_R^i\to Z'\phi$ annihilation. If $\lambda_{\phi s}$ is too small, the FOPT strength is so strong that the entropy production during reheating dilutes the BAU to an unacceptable low value, as covered by the orange shaded region in the figure. We have checked that, without the FOPT, the same parameter space in Fig.~\ref{fig:leptogenesis} cannot realize a conventional thermal leptogenesis in the $B-L$ model, which typically requires a CP asymmetry $\mO(30)$ larger than the Davidson-Ibarra bound due to the large thermal washout effects from processes involving $Z'$ and $\phi$. Therefore, our model has opened up new parameter space for a novel kind of leptogenesis.

In this scenario, the relevant energy scale is about $10^{11}$ GeV, which is not accessible at any current or near-future colliders. However, the GWs as byproducts of the $U(1)_{B-L}$ breaking may help to probe the scenario, although those signals could not serve as smoking guns for this specific mechanism. Thus, we briefly comment on the possible signals. In our scenario, there are two sources of the GWs: first, the $U(1)_{B-L}$ FOPT itself generates GWs via vacuum bubble collision, sound waves and magneto-hydrodynamics (MHD) turbulence in the plasma~\cite{Jinno:2016knw,Iso:2017uuu,Marzo:2018nov,Ellis:2019oqb,Bian:2019szo,Ellis:2020nnr}; second, the cosmic strings forming after the $U(1)_{B-L}$ breaking keep emitting GWs during the evolution of the Universe~\cite{Buchmuller:2013lra,Dror:2019syi,Fornal:2020esl,Samanta:2020cdk,Masoud:2021prr,Bian:2021vmi,Buchmuller:2021mbb}.

As an illustration, we adopt $\lambda_{\phi s}=3.5$ and $g_{B-L}=0.05$ as a benchmark (shown as the green star in Fig.~\ref{fig:leptogenesis}) to calculate the GW spectrum today after the cosmological redshift. For the FOPT GWs, $T_p=6.1\times10^{10}$ GeV, and the energy budget depends on the evolution of the wall velocity, thus we tried both schemes from Ref.~\cite{Hoche:2020ysm} (with $\mP_{1\to N}\propto\gamma_w^2$) and Ref.~\cite{Gouttenoire:2021kjv} (with $\mP_{1\to N}\propto\gamma_w$). For the former case, as the friction increases rapidly with $\gamma_w$, the bubble walls have reached the terminal velocity at $T_p$, thus the sound wave and MHD contributions dominate~\cite{Ellis:2018mja}, and we make use of the efficiency factor $\kappa_V$ derived in Ref.~\cite{Espinosa:2010hh}; while for the latter case, the walls are still accelerating at $T_p$, and hence the bubble collision contribution dominates, and we adopt the method in Refs.~\cite{Ellis:2019oqb,Ellis:2020nnr} to obtain the efficiency factor $\kappa_{\rm col}$. With the efficiency coefficients in hand, the FOPT GW spectra are evaluated using the numerical formulae in Refs.~\cite{Caprini:2015zlo,Caprini:2019egz}.\footnote{In the sound wave dominant case, the extra suppression factor from the finite duration of sound wave period is taken into account~\cite{Ellis:2018mja,Ellis:2019oqb,Guo:2020grp}.} For the cosmic strings GWs, the spectrum is determined by the dimensionless combination $G\mu$, where $G=1/M_{\rm Pl}^2$ is the Newton's constant of gravitation, and $\mu\sim v_\phi^2$ is the tension of the strings. For our benchmark, $G\mu\approx10^{-14}$, and we use the numerical results in Refs.~\cite{Auclair:2019wcv,Blanco-Pillado:2017oxo,Binetruy:2012ze,Blanco-Pillado:2013qja} to derive the GW spectrum.\footnote{See Refs.~\cite{Bian:2022tju,Zhao:2022cnn} for recent research on cosmic string GW simulations and experimental constraints.}

\begin{figure}
\centering
\includegraphics[scale=0.55]{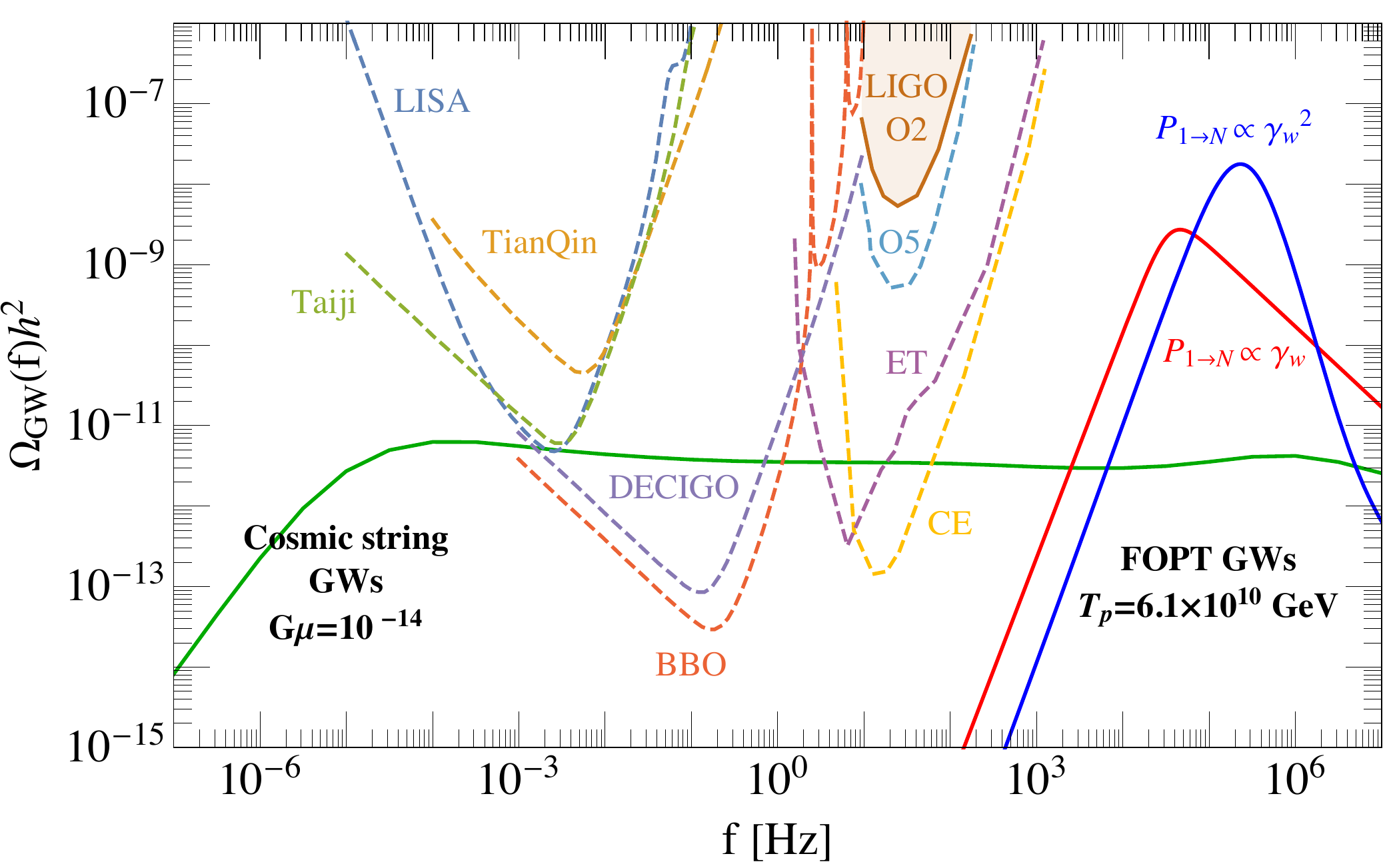}
\caption{The GW spectra for the benchmark $\lambda_{\phi s}=3.5$ and $g_{B-L}=0.05$ (marked as green star in Fig.~\ref{fig:leptogenesis}). Both the spectra from cosmic strings and FOPT are shown, and in the latter case both two possibilities of $\mP_{1\to N}\propto\gamma_w^2$ (sound wave dominant) and $\mP_{1\to N}\propto\gamma_w$ (bubble collision dominant) are considered.}
\label{fig:gws}
\end{figure}

The GW spectra for our benchmark point are given in Fig.~\ref{fig:gws}, where the expected sensitivity curves for the space-based laser interferometers LISA~\cite{LISA:2017pwj}, TianQin~\cite{TianQin:2015yph,Hu:2017yoc,TianQin:2020hid}, Taiji~\cite{Hu:2017mde,Ruan:2018tsw}, BBO~\cite{Crowder:2005nr} and DECIGO~\cite{Kawamura:2011zz}, and the ground-based interferometers LIGO~\cite{LIGOScientific:2014qfs,LIGOScientific:2019vic}, CE~\cite{Reitze:2019iox} and ET~\cite{Punturo:2010zz,Hild:2010id,Sathyaprakash:2012jk} are also shown. We first see that the FOPT GWs spectra peak at $\sim10^5$ Hz, which is too high to be detected by the near-future instruments. For heavier RHNs and hence higher FOPT scales, the typical peak frequency is even higher and hence more difficult to probe. However, the cosmic string GW spectrum is rather flat and could be reached by quite a few future detectors such as BBO, DECIGO, CE and ET. For heavier RHNs, $G\mu$ is larger, and the signal strength becomes stronger that LISA, TianQin and Taiji can also probe the scenario. Therefore, we conclude that the cosmic strings induced GWs are hopeful to be seen at the future detectors, although this is a general feature of all the high-scale $U(1)$ breaking new physics models, not specifically for our extended $B-L$ model.

\section{Conclusion}\label{sec:conclusion}

In this article, we apply the mechanism of baryogenesis induced by ultra-relativistic bubble walls to the leptogenesis case. After giving a general discussion on the dynamics of such a scenario, we build an extended $B-L$ model to demonstrate the viable parameter space realizing the mechanism. We have shown that the mechanism requires a trade-off between the strength of FOPT and the level of reheating, and the successful FOPT leptogenesis requires RHN mass $\gtrsim10^{11}$ GeV assuming the Davidson-Ibarra bound. Meanwhile, we verify that the same parameter space cannot generate sufficient BAU within the conventional thermal leptogenesis mechanism. Therefore, our research provides a novel approach to realize leptogenesis. While the frequency of GW signals from FOPT is too high to be probed at the detectors, the GWs emitted by the cosmic strings from $U(1)_{B-L}$ breaking might be seen at the near-future detectors such as LISA, TianQin, Taiji, CE and ET.\\

\noindent{\bf Note added.} Soon after the completion of this manuscript, Ref.~\cite{Dasgupta:2022isg} appears, which applies the same mechanism to the minimal classically conformal $B-L$ model in the resonant leptogenesis regime.

\acknowledgments

We would like to thank Iason Baldes, Ville Vaskonen and Shao-Jiang Wang for the very useful and inspiring discussions. This work is supported by the National Science Foundation under grant number PHY-1820891, and PHY-2112680, and the University of Nebraska Foundation.

\bibliographystyle{JHEP-2-2.bst}
\bibliography{references}

\end{document}